# The bump-on-tail instability excited by energetic electrons in helicon plasma

Shi-Jie Zhang[1#], Dong Jing[2,3#], Lei Chang[1*], Kai-Jun Fu[1], Chao Wang[1], Zi-Chen Kan[1], Ye Tao[1], Jing-Jing Ma[1], Ji-Kai Sun[1], Ding-Zhou Li[1], Ilya Zadiriev[4], Elena Kralkina[4], Shin-Jae You[5]

[1]State Key Laboratory of Power Transmission Equipment Technology, School of Electrical Engineering, Chongqing University, Chongqing, 400044, China

[2]Institute of Plasma Physics, HFIPS, Chinese Academy of Sciences, Hefei 230031, China

[3]University of Science and Technology of China, Hefei, 230026, China

[4]Physical Electronics Department, Faculty of Physics, Lomonosov Moscow State University, GSP-1, Leninskie Gory, Moscow, 119991, Russian Federation

[5]Department of Physics, Chungnam National University, Daejeon, 34134, Republic of Korea

[#]These two authors contribute equally to the paper.

[*]Correspondence Email: leichang@cqu.edu.cn (Lei Chang)

**Abstract**

This work explores for the first time bump-on-tail (BOT) instability excited by energetic electrons in helicon plasma. The Berk-Breizman model that developed for the wave-particle interaction and resulted instability in magnetic fusion is used. Details of the BOT instability are computed referring to typical helicon discharge conditions. Parameter studies are also conducted to reveal the effects of collisionality and energetic drive, to account for high-pressure and high-power senarios respectively. It is found that under the HXHM (high magnetic field helicon experiment) experimental parameters, the disturbed distribution function oscillates explosively at the initial stage of BOT instability excitation, and the wave frequency shift does not appear, i.e., the steady-state solution always exists under this mode. In the process of restoring stability, the exchange of energetic particles and wave energy is concurrent with the change of wave amplitude. As the Krook operator increases (i.e., from 0.1 to 1), the saturation level of the electric field and the instability enhance. Additionally, there have a bigger disturbance for the initial EEDF (electron energy distribution function) in high-power helicon devices, so that the energy exchange between waves and energetic particles is stronger as well. Moreover, BOT instability effects the density and flux of bulk plasma, and the flux increases with the Krook operator. The effect of BOT instability is one order of magnitude larger on rotating plasma than that on stationary plasma. These findings present a full picture of BOT instability in helicon plasma and are valuable to controlling it for efficient and safe applications, e.g., high-power space plasma propulsion and plasma material interactions using helicon source.

# 1. Introduction

Helicon plasma has been applied in several areas, e.g. material deposition, etching for semiconductors [1], plasma thrusters [2, 3] and nuclear fusion [4], due to its high ionization rate and high plasma density, since the first helicon discharge achieved by R. D. Boswell in 1970s [5]. There have been plenty of advances in experiments and theories, such as the studies of antenna types [6], magnetic field configurations [7], helicon wave and TG modes [8, 9], power deposition [10], blue-core discharge [11] as well as turbulence [12]. The role of energetic electrons attracts great attention among the studies mentioned above. F. F. Chen was the first to propose that the Landau damping mechanism, which is a process involving energetic electrons with velocities resonant to the wave phase, could be responsible for highly efficient ionization [13]. Subsequently, the energetic electrons have been detected by A.R. Ellingboe et al. using RF-modulated emission of $Ar^+$ light [14], A.W. Molvik et al. using retarding-potential analyzer (RPA)[15], and Chen et al. using Langmuir probes [16]. Recently, energetic electrons have also been observed in downstream magnetic nozzles where these energetic electrons are only considered from upstream helicon sources [17, 18]. Meanwhile, it is found that energetic electrons do not appear uniformly in radius according to the electron energy probability function (EEPF) measured in Chi-Kung experimental setup [19]. One of the predictable influences of energetic electrons is ionization. Large amplitude standing helicon wave under the antenna is responsible for populating the electron energy distribution function (EEDF) which can interact resonantly with the potential well of the traveling helicon wave, and thus produce energetic electrons [14]. The energetic electrons which have energy corresponding to a large ionization cross section will predominantly contribute to ionization [20]. In addition, energetic electrons would lead to the excitation of ion-acoustic parametric turbulence [21, 22], appearance of large electric potential of double layer [23], as well as effect the drift wave instability according to the flowing plasma model [24]. As energetic particles existing in bulk plasma, the beam-plasma instability should be important. There has been an estimate of the quasilinear growth rate for energetic electron beam in a relatively cold plasma, i.e. in order of $3\times10^8$ s$^{-1}$. This corresponds to a distance along the beam of order 3 cm and beamwidth $\delta v/v \sim 1/5$ [20]. However, it is rarely reported the quantitative effect of instabilities driven by energetic electrons. In this paper, the evolutions of bump-on-tail (BOT) instability driven by energetic electrons and their effects on bulk helicon plasma are evaluated for the first time.

Energetic particles that indicated by the bumped tail of Maxwellian distribution could generate instability in bulk plasma when the resonant condition is satisfied. This topic has been exceedingly investigated in magnetic nuclear fusion, and Berk-Breizman (BB) model [25, 30-36] is one of the most elegant models to describe the underlying physics. Inspired by the natural similarities of magnetized plasmas, the BOT problem of energetic electrons in helicon plasma will be studied by BB model in this work. The results show that there is indeed BOT instability of growth rate $1.71\times10^8$ s$^{-1}$ caused by energetic electrons existing in the high magnetic field helicon experiment (HMHX), in which the BOT feature was measured in detail [26]. The disturbed velocity distribution of electrons is 2.5% and the excited electric field evolves to saturation. We shall show that the BOT instability will lead to pondermotive force as well as plasma flux change. Moreover, although it is hard to measure the nonlinear kinetic effects caused by the BOT instability under the antenna, fast electrons oscillating in standing waves under the antenna are subject to beam-plasma instability and may be thermalized rapidly [27]. Particularly, it is notable that changes in the axial magnetic field or RF power could cause order of magnitude changes in the energetic electrons current [15]. Therefore, it is necessary to investigate the influence of BOT instability especially under the antenna and the higher-power helicon plasmas [28]. In these circumstances, energetic electrons should be a large population in the bulk plasma, so that the BOT instability will cause severe problems in terms of power coupling, steady operation, thermal load, and reliable usage.

The paper is organized as follows: Sec. 2 describes the theoretical model, Sec. 3 shows the numerical scheme, Sec. 4 presents the calculated BOT instability referring to the HMHX device, higher-power helicon discharge senarios, and its influences on the bulk plasma, respectively, and Sec. 5 concludes the paper and gives remarks for future research.

## 2. Theoretical model

The physical essence of BOT problem is a resonant interaction of energetic particles with weakly unstable discrete modes for which the linear growth rate is much less than the mode frequency [29]. The nonlinear evolution of this kinetic system is maintained by the balance between source and relaxation processes. The BB model is an elegant and extensively verified perturbation approach for the BOT problem. In the BB model, the properties of perturbing modes are determined robustly by bulk plasma while energetic particles only effect their amplitude. It has been surprisingly successful in terms of explaining many complex phenomena observed in magnetic fusion experiments [37-39], including frequency splitting and chirping. The present study makes use of this elegant and verified theory to treat the BOT problem in helicon plasma, which to our best knowledge is the first time. While details of the BB model could be found in previous publications [25, 40], key aspects about the model are summarized below for easy reference.

The one-dimensional Boltzmann equation for the distribution function of energetic electrons is:

$$\frac{\partial F}{\partial t} + v \frac{\partial F}{\partial x} + \frac{e}{m} \hat{E}(t) \times \cos(kx - \omega t) \frac{\partial F}{\partial v} = -\nu(F - F_0), \qquad (1)$$

with $\nu$ the annihilation rate, labelling Krook-type relaxation process. Here, the mass and charge of electron are denoted by $m$ and $e$, respectively. The energy conservation ensures that the time rate of the change of wave energy $\partial WE/\partial t$ equals to the negative power dissipated into the background plasma, i.e. $-2\gamma_d WE$, plus the power transferred from energetic particles to waves,

$$P = -\frac{e\omega}{k} \int dxdv E(x,t) F(x,v,t). \tag{2}$$

The wave energy of plasmas takes into account the wave energy and the kinetic energy generated by the plasma frequency oscillation, where $WE = \int dx E^2(x,t)/4\pi$ is integrated at one wavelength. By analyzing these equations, the following results are derived,

$$\frac{\partial \hat{E}(t)}{\partial t} = -\frac{4\pi e\omega}{k} \text{Re} \int f_1 dv - \gamma_d \hat{E}(t), \tag{3}$$

where $\gamma_d$ means the intrinsic damping rate from background plasma. Its expression is $\gamma_d \equiv \nu_{ei}/2$, where $\nu_{ei}$ is the collision frequency between energetic electrons and the background plasma. The physical quantities have been represented as Fourier series for further theoretical analysis

$$F = F_0 + f_0 + \sum_{n=1}^{\infty} [f_n \exp(in\psi) + c.c], \tag{4}$$

where $\psi = kx - \omega t$. In other words, this shows a traveling wave solution which has a spatial period denoted by $\lambda = 2\pi/k$ and a carrier frequency denoted by $\omega$. If $F$ is truncated at $n = 2$, Eq. (1) for $n = 0, 1, 2$ then become:

$$\frac{\partial f_0}{\partial t} + \nu f_0 = -\frac{\omega_B^2}{2} \frac{\partial (f_1 + f_1^*)}{\partial u}, \tag{5a}$$

$$\frac{\partial f_1}{\partial t} + iuf_1 + \nu f_1 = -\frac{\omega_B^2}{2} \frac{\partial (F_0 + f_0 + f_2)}{\partial u}, \tag{5b}$$

$$\frac{\partial f_2}{\partial t} + 2iuf_2 + \nu f_2 = -\frac{\omega_B^2}{2} \frac{\partial f_1}{\partial u} + \mathcal{O}(\omega_B^2 f_3), \tag{5c}$$

where initial distribution $f_0$ is independent on $x$. In Eq. (5), the electric field is normalized in terms of $\omega_B^2 = ek\hat{E}(t)/m$. Now, $\int f_1 dv$ needed for solving Eq. (3) can be obtained by the integration of Eq. (5). The final equation that represents the BB model is

$$\frac{d}{dt}\omega_B^2 = (\gamma_L - \gamma_d)\omega_B^2(t) - \frac{\gamma_L}{2} \int_{\frac{t}{2}}^{t} dt' (t-t')^2 \omega_B^2(t') \times \\ \int_{t-t'}^{t'} dt_1 \exp[-\nu(2t - t' - t_1)] \times \omega_B^2(t_1) \omega_B^2(t' + t_1 - t), \tag{6}$$

where the wave growth rate is

$$\gamma_L = 2\pi^2 (e^2\omega/mk^2) \frac{\partial F_0(\omega/k)}{\partial v}, \tag{7}$$

and $v = \omega/k$ is the velocity of particles trapped in the wave field. The slope of $F_0$, assumed as a positive constant $\partial F_0/\partial v > 0$ within narrow range of resonant velocity, determines the linear wave growth rate $\gamma_L$. Normalization can further lead to

$$\frac{dA}{d\tau} = A(\tau) - \frac{1}{2} \int_0^{\frac{\tau}{2}} dz\, z^2 A(\tau - z) \times \int_0^{\tau - 2z} dx \exp[-\hat{\nu}(2z + \tag{8}$$

$$x)] \times A(\tau - z - x)A(\tau - 2z - x),$$

$A$, $\tau$ and $\hat{v}$ are respectively,

$$\begin{cases} A = \left[\dfrac{\omega_B^2}{(\gamma_L - \gamma_d)^2}\right]\left[\dfrac{\gamma_L}{\gamma_L - \gamma_d}\right]^{\frac{1}{2}}, \\ \tau = \dfrac{\gamma_L - \gamma_d}{t}, \\ \hat{v} = \dfrac{v}{\gamma_L - \gamma_d}. \end{cases} \quad (9)$$

In the reference [25], the marginal instability with $\gamma_L - \gamma_d \ll \gamma_L$ will evolve to steady state $A_0 = 2\sqrt{2}\hat{v}^2$ from Eq. (8) for $\tau \to \infty$ and the critical threshold corresponding to the occurrence of instability will be $\hat{v} > \hat{v}_{cr} = 4.38$. However, it is different from our study presented here (i.e. $\gamma_d < \gamma_L$), the instability will eventually evolve into a steady state for relatively small value of $\hat{v}$, such as $\hat{v} = 0.019$. Nevertheless, we shall show that the BB model is still applicable and various interesting physical phenomenon will be given.

## 3. Numerical scheme

Benefitting from the successful work of Matthew Lilley and co-workers [36, 41-45], we shall make use of their code to solve Eq. (8) in the following sections. The code is also named in phrase of BOT and developed particularly to treat the bump-on-tail problem and solve the BB model. While details about the code can be found in the website [50] and publications mentioned there, key points are summaried below.

The BOT code use the normalised variables $C = \omega_B^2 / \gamma_L^2$, $G = (2\pi|e|^2 \omega_{pe} / (m\gamma_L^2 k))F$, $\tau = \gamma_L t$, $\hat{\gamma}_d = \gamma_d / \gamma_L$, $\Omega = (kv - \omega_{pe})/\gamma_L$, $\bar{v} = v/\gamma_L$, and similarly for $\bar{\alpha}$ and $\bar{\beta}$, transforming Eq. (3) and Eq. (5) into a set of advection equations:

$$\dfrac{\partial \chi_n}{\partial \tau} - n\dfrac{\partial \chi_n}{\partial s} + \left[\bar{v}^3 s^2 - i\bar{\alpha}^2 s + \bar{\beta}\right]G_n = R_n(s,\tau) + \delta_{1,n}\dfrac{1}{\sqrt{2\pi}}C\delta(s), \quad (10)$$

$$\dfrac{\partial C}{\partial \tau} + \hat{\gamma}_d C = 2\sqrt{2\pi}G_1(0,\tau), \quad (11)$$

where $\chi_n = \dfrac{1}{\sqrt{2\pi}}\int_{-\infty}^{+\infty} G_n \exp(-i\Omega s)d\Omega$ is the Fourier transform of $G_n$, $\delta_{1,n}$ is the Kronecker delta and $R_n(s,\tau)$ represents the non-linearity in the equations. Differential collision operators are displaced by algebraic operators. The equations are integrated using the method of characteristics, concretely, changing variables to $s' = s + n\tau$, $\tau' = \tau$ and integrating in $\tau'$ by the trapezoid rule.

This methodology generates a system of integral equations that inherently circumvent the Courant-Friedrichs-Lewy (CFL) stability condition—a constraint typically imposed by explicit finite difference methods.

In this paper, we investigate a weakly nonlinear instability characterized by the condition $\omega_B / \gamma_L \ll 1$. Within this regime, the nonlinear effects perturb a localized region of phase space near the resonance. Consequently, a narrow velocity range in the simulation is set. In Fourier space, this corresponds to lower resolution and the smaller number of conjugate velocity points. The BOT code is numerically stable as long as $|C|\Delta\tau < 2/|s|_{max}$, with $C = \omega_B^2 / \gamma_L^2$. In this article, $|s|_{max}$ is set to 10 and all the calculations are numerically stable.

## 4. Results and discussion

In this section, we choose a well-developed helicon device for modelling and computation, and utilized the experimental data in HMHX to calculate the growth rate and dissipation of the BOT instability. Moreover, we explore the effects of Krook collision on the BOT instability, the effects of $\gamma_L$ on the BOT instability and the effects of BOT instability on the background plasma.

### 4.1. BOT instability in HMHX experiment

HMHX is an advanced helicon discharge device, as shown in Fig. 1, and it is mainly used for plasma wall interaction (PWI) with high magnetic field ($B_0 > 5000$ G) and high plasma density ($n_e > 10^{19}$ m$^{-3}$). The experimental setup consists of two main parts: an helicon wave plasma source region and a PWI region, and the internal antenna is an 18 cm-long, 3.5 cm-diameter, and right-handed Nagoya III helicon antenna. In the experiment, an RF-compensated Langmuir probe is placed at 0.17 m downstream of the HWP source region interface to measure the spatial plasma parameters in the PWI region.[26]

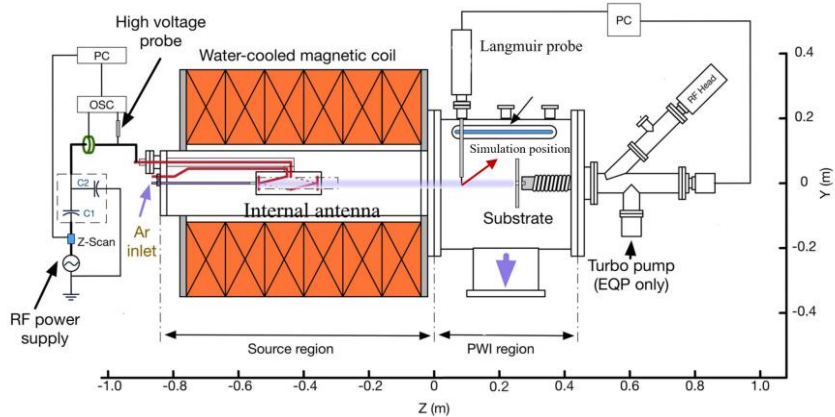

**Figure 1** Schematic of HMHX [26]. The red arrow demonstrates the simulation position.

We first calculate the input conditions of BOT from the HMHX parameters. The EEPF exhibits a BOT feature at helicon discharge conditions of power 1500 W and magnetic field 1900 G. In order to obtain one-dimensional velocity distribution function, we divide the EEPF plot by $\sqrt{3}$ due to $v^2 = v_x^2 + v_y^2 + v_z^2$ and treat the abscissa as the electron kinetic energy. The experimental distribution of electron velocity is measured and showed in Fig. 2. The inset figure shows an almost linear positive slope in the energetic electron distribution. Since the velocity range of interest is significantly narrower than the overall width of the distribution function $F_0$, the slope can be approximated as constant[41], with its average value denoted as $\partial f_0/\partial v$ =622.38 m$^{-5}$s$^2$. We then set plasma frequency to be the frequency of perturbing mode as done in previous study [41], i.e. $\omega = \omega_{pe} = \sqrt{n_0 e^2/\epsilon_0 m_e} = 1.78 \times 10^9$ s$^{-1}$. If taking resonant velocity as $v = 3.134 \times 10^6$ m/s (displayed by red dot), with the resonant condition of $\omega = kv$, we can have the growth rate of wave mode, i.e. $\gamma_L$ =1.71×10$^8$ s$^{-1}$. In that case, the wave growth rate is an order of magnitude less than the wave frequency, which is consistent with presented energetic-particle driven experiments, where $\gamma_L/\omega$ is estimated within the range of 10%–30% [46, 47]. When the mode is excited, the damping of wave will be mainly electron-electron and electron-ion collision in the bulk plasma, $\nu_{ee}$ =3.285×10$^7$ s$^{-1}$ and $\nu_{ei}$ =1.64×10$^7$ s$^{-1}$, respectively. Therefore, the damping rate in BOT model is $\gamma_d = (\nu_{ee} + \nu_{ei})/2 = 2.4625 \times 10^7$ s$^{-1}$ and then $\gamma_d/\gamma_L$ = 0.1437. In the paper, Krook-type collision operator is considered as relaxation process, that is, the collision between energetic electron and bulk "cold" electron $\nu = 1/\tau_c$, where the averaged collision time is

$$\tau_c = \frac{3\sqrt{2m_e}\pi^{\frac{3}{2}}\varepsilon_0^2}{n_0 e^4 \ln\Lambda} (T_{bulk} + T_{Ep})^{3/2}. \qquad (12)$$

In BOT code, the dimensionless number of the Krook operator $\beta/(\gamma_L - \gamma_d)$ is set to 0.019 (corresponding to Tab. 1).

All parameters about HMHX plasma and the BOT code inputs are listed in Tab. 1.

**Table 1**. The parameters of HMHX (inputs for BOT code).

| Name of Parameter | Value | Name of Parameter | Value |
| --- | --- | --- | --- |
| Plasma density (m$^{-3}$) | 1×10$^{19}$ | Electron temperature (eV) | 6 |
| Plasma frequency (s$^{-1}$) | 1.78×10$^9$ | $\nu_{ee}$ (s$^{-1}$) | 3.285×10$^7$ |
| Debye length (m) | 5.76×10$^{-6}$ | $\nu_{ei}$ (s$^{-1}$) | 1.64×10$^7$ |
| The slope at resonance (m$^{-5}$s$^2$) | 622.38 | Initial bounce frequency $\omega_B$ in units of $\gamma_L$ | 1×10$^{-3}$ |
| Resonance velocity $v$ (m/s) | 3.134×10$^6$ | Wave damping rate $\gamma_d$ (s$^{-1}$) | 2.46×10$^7$ |
| $\gamma_d$ in units of $\gamma_L$ | 0.1437 | Wave growth rate $\gamma_L$ (s$^{-1}$) | 1.71×10$^8$ |
| Krook operator $\beta/(\gamma_L - \gamma_d)$ | 0.019 | | |

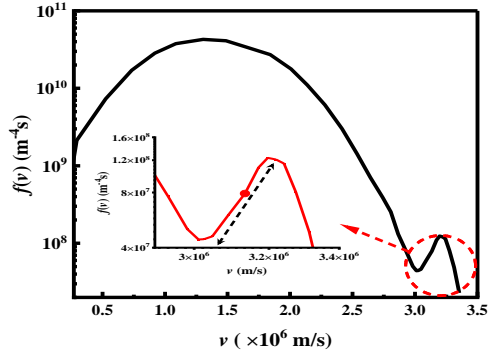

**Figure 2**. The distribution function of electron velocity in HMHX device[26]. The inset figure is the zoomed distribution of energetic electron, showing a clear BOT feature.

Then, we compute the time evolution of the BOT instability, utilizing the HMHX parameters as inputs for the BOT code. The temporal evolution of $\omega_B^2 = ekE_1/m_e$, i.e. electric field of the instability is shown in Fig. 3, where all physical quantities have been normalized. At first, the instability is exponentially excited due to inverse Landau damping with the resonant condition of $v_{\text{phase}} = v_{\text{particle}}$.

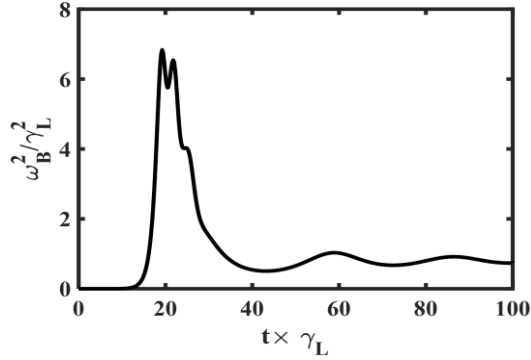

**Figure 3**. Evolution of BOT instability with parameters in HMHX calculated by the BOT code: the evolution of electric field.

To show more details, we plot the evolution of distribution functions at the resonance velocity. At $t \times \gamma_L \approx 18$, the initial distribution at resonance presenting in Fig. 4 is flattened, and the growth of BOT instability stops and subsequently decreases because of collisional damping. Finally, the energy transferred to wave from energetic electrons balances with the wave energy dissipated into the bulk plasma. From Fig. 4, the slope of $F$ is almost the same at $t \times \gamma_L \approx 18$ and 100. The change of EEDF is around 2.5%.

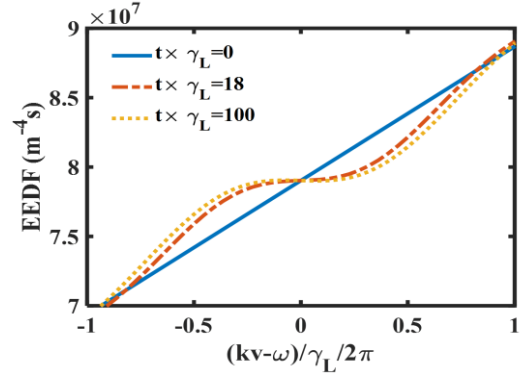

**Figure 4**. Evolution of BOT instability with parameters in HMHX calculated by the BOT code: the distributions of energetic electrons at three temporal moments, where $t \times \gamma_L = 0$ replaces initial distribution and x-axis is normalized (i.e. Min-Max Normalization) velocity.

Figure 5 displays the 2D evolution of distribution function. In Fig. 5, there are small perturbations around the resonance at the beginning of interaction, i.e. $t \times \gamma_L \approx 18$, however, those small perturbations gradually disappear over time. It is relaxation that restores the perturbances to equilibrium.

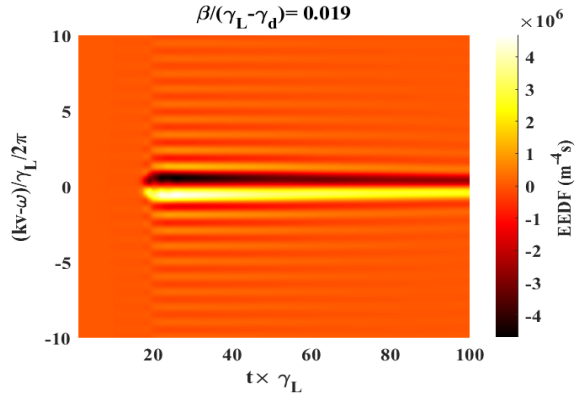

**Figure 5**. Evolution of 2D disturbed distribution with Krook coefficient $\beta/(\gamma_L - \gamma_d) = 0.019$ calculated by the BOT code.

Next, we study the disturbance part of EEDF ($f_0$) and its phase space, which gives the core features of BOT instability. As shown in Fig. 6, the velocity distribution function $f_0$ changes with time is observed. It can be seen that in Fig. 6($a_1$-$d_1$) the number of energetic electrons near the right side of the resonance velocity decreases, while the number on the left side increases. This means that some of the energetic eletrons transfer their energy to the waves. By comparing Fig. 6($a_1$) with Fig. 6($a_2$), we observe that the electron distribution at the resonance point is disturbed when the instability is excited. At this moment, the wave field captures electrons near the resonance region and forms phase islands. When $t \times \gamma_L = 36$ (corresponding to Figs. 6($b_1$) and ($b_2$)), the perturbation in the distribution function intensifies further, with electrons even far from the resonance region being significantly perturbed. This results in the disappearance of phase islands and the onset of 'phase mixing'. Fig. 6(c) shows a transition process to steady state. At $t \times \gamma_L = 500$ (i.e. Fig. 6(d$_1$) and 6(d$_2$)), the evolution of the distribution function nearly stabilizes. This indicates that the energy transferred from energetic electrons to the waves is balanced by the wave energy dissipated into the background plasma, while the slope of the resonance region in the distribution function becomes flattened.

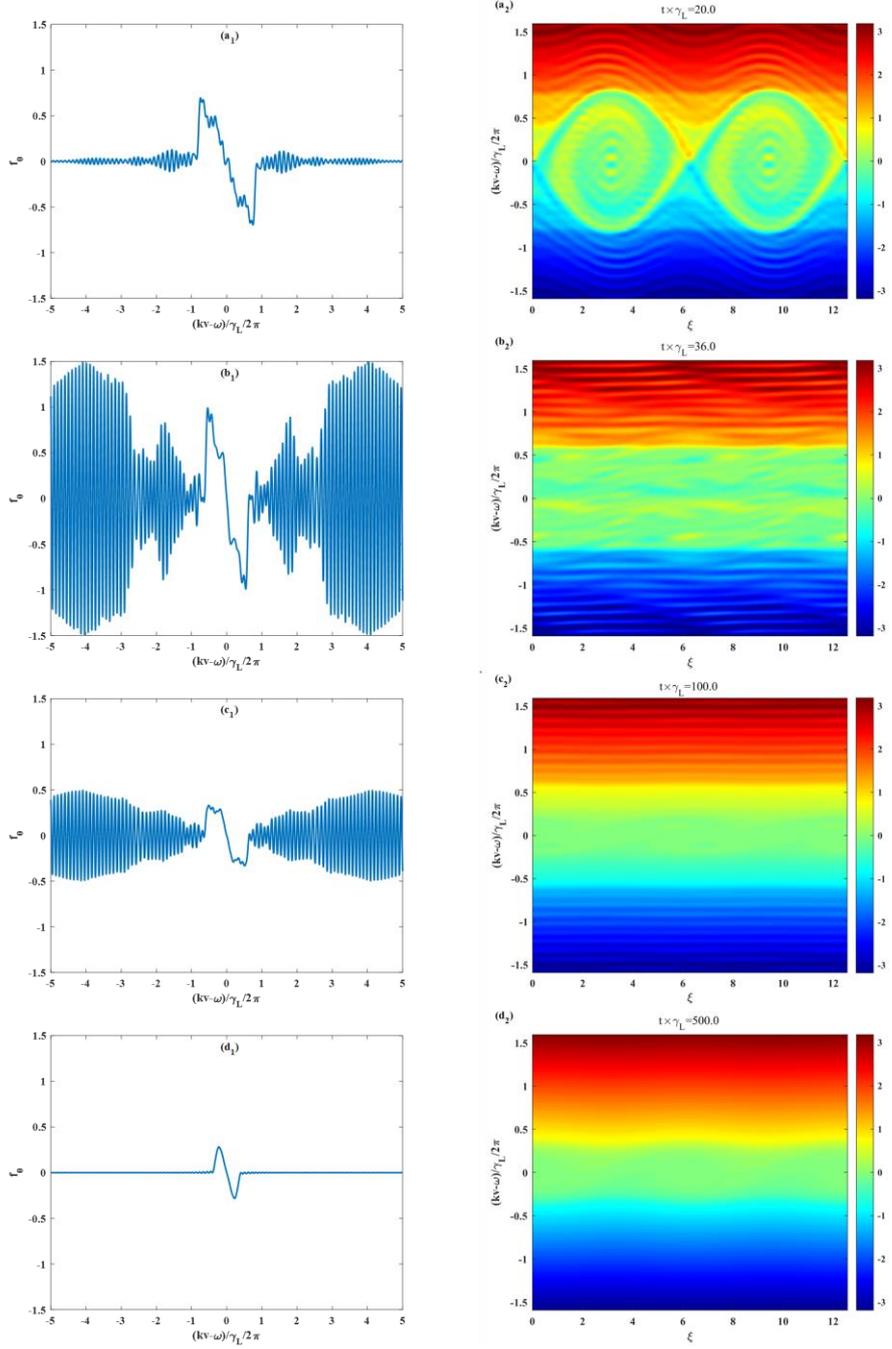

**Figure 6**. The temporal evolution of the velocity distribution function and phase space dynamics, where (a-d) represents $t \times \gamma_L =$ 20, 36, 100, 500, respectively. Other conditions are $\gamma_d / \gamma_L = 0.1437$ and $\beta(\gamma_L - \gamma_d) = 0.019$.

Moreover, the BOT instability may induce frequency sweeping; therefore, its temporal frequency variations merit detailed investigation. Figure 7 illustrates the wave spectrum evolution of the instability. The frequency exhibits no temporal variation, maintaining a near-constant value of $\omega/\gamma_L/2\pi = 10$ throughout the entire time domain. The observed frequency stabilization can be explained by resonant energy exchange mechanisms. Specifically, particles with velocities marginally exceeding the phase velocity contribute energy to the wave (driving growth), while those slightly below the phase velocity extract energy from the wave (inducing damping). This self-regulating energy transfer predominantly modulates wave amplitude while maintaining spectral stability, consistent with the steady-state solution in the system.

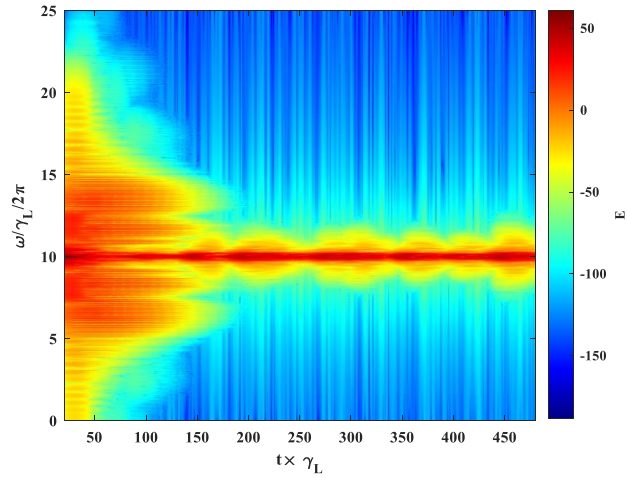

**Figure 7**. A full picture of the wave spectrum of BOT instability.

Futhermore, we explore the power coupling features during the BOT instability, because the energy exchange between wave and energetic electrons can effect the power coupling of plasma. One can see from Fig. 8(a) that the profile of power transferred by the energetic electrons to the wave is the same to that of electric field. This profile reflects the wave energy obtained from electron kinetic energy. If only disturbed EEDF is considered, not all of power transfer to wave, like in Fig. 8(b). The result of EEDF indicates that the BB model featured by perturbing method is properly used for the BOT instability in HMHX helicon plasma. The similar instability with Krook relaxation process has been well investigated in a previous study [48].

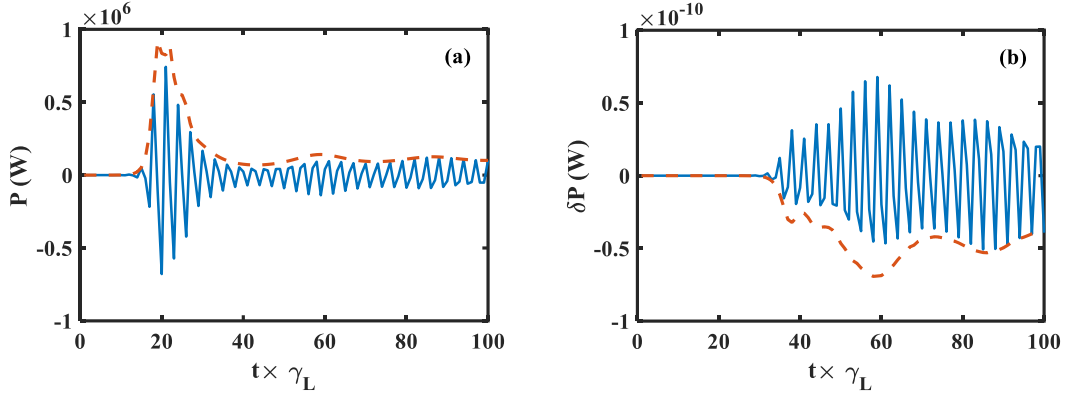

**Figure 8**. The total power transferred by the energetic electrons to the wave (a) and the power transferred by the change of EEDF (b), where dotted line represent overall amplitude.

Finally, we investigate the effects of the BOT instability on the temperature of energetic electrons. We introduce $\delta T$ for perturbed temperature:

$$\delta T = \frac{1}{2}m\frac{\int v^2 f dv}{\int f dv} - T_0, \qquad (13)$$

where $T_0$ is the temperature of the undisturbed energetic electrons. In HMHX, $T_0$ is 28.9 eV. The results of $\delta T$ and corresponding $\omega_B^2/\gamma_L^2$ are shown in Fig. 9. It shows that the BOT instability causes negative $\delta T$ value due to energy conservation. The most significant change appears at the same time of $t \times \gamma_L \approx 18$ that corresponds to the maximum electric field from Fig.9. The $\delta T$ initially decreases sharply, consistent with the exponential growth of electric field, indicating strong wave-particle interactions involving energetic electrons. After $t \times \gamma_L \approx 40$, the electric field exhibits periodic modulation, while $\delta T$ resumes its upward trend. And it can be predicted that $\delta T$ will reach saturation near $-5 \times 10^4$ eV.

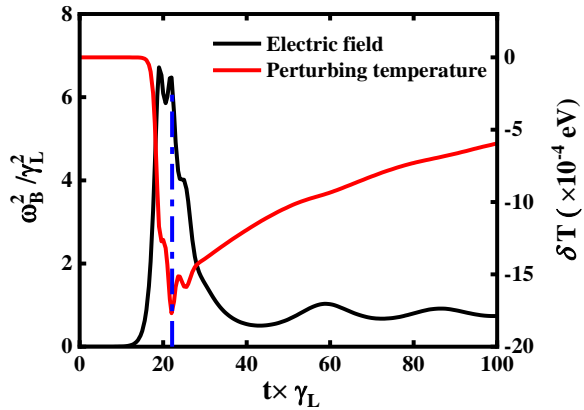

**Figure 9**. The variation of energetic electron temperature for parameters in HMHX during the BOT instability, where the electric field is the same as in Fig. 3 and blue dotted line is at $t \times \gamma_L =22$.

### 4.2. Effects of Krook operator on BOT instability

In the BOT instability analyzed above, the main relaxation process is regarded as Krook collision, and the population of energetic electrons is very small in the HMHX experiment. Now, we consider the case of bigger population of energetic electrons through studying the effect of Krook operator on the BOT instability. The employed EEDF is shown by red line in Fig. 10, in which the EEDF in Fig. 2 is also plotted in black dashed line for comparison.

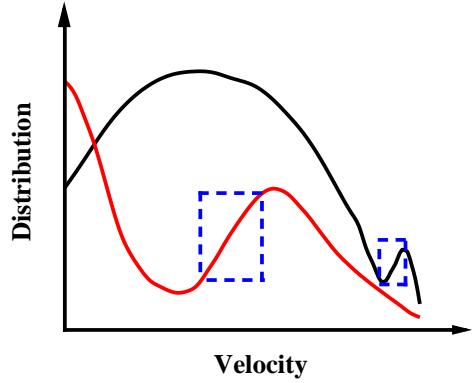

**Figure 10**. Two types of distribution: red, a distribution with more energetic electrons; black, the distribution in HMHX, where blue dotted boxes indicate the area of appearance for BOT instability.

From the expressions of linear growth rate $\gamma_L$ and Krook collision $\beta$, the ratio that determining the intensity of BOT instability can be written as,

$$\beta/\gamma_L \approx \frac{n^{3/2}}{(T_b+T_{ep})^{3/2}\frac{\partial F}{\partial v}v^2}. \tag{14}$$

It means that the BOT instability would be stronger with higher population of energetic electrons. On the basis of Eq. (7), the value of $\gamma_L$ is related to the slope of EEDF near the resonance velocity. We assume that the slope of the distribution function of the energetic electron population at resonance is invariant and change Krook collision $\beta$ to simulate the BOT instability with higher energetic electron population.

Figure 11 demonstrates the dependence of BOT instability on Krook-type collisional relaxation. The excitation of BOT instability is governed by the positive slope at the resonance point, while its subsequent stabilization relys on collisional relaxation. As shown in Figs. 11(a-d), higher values of $\beta/(\gamma_L - \gamma_d)$ correlate with enhanced electric field saturation amplitudes. This trend is more obvious in Fig. 13(a), confirming the critical role of collisionality in regulating nonlinear saturation. The changes of temperature in Fig. 12 resemble overturned electric field. For $\beta/(\gamma_L - \gamma_d) = 0.019$ (Fig. 9), the temperature saturation exhibits a delayed response compared to the electric field. In contrast, when $\beta/(\gamma_L - \gamma_d) = 1$, temperature and electric field saturate synchronously, as evidenced by the comparative analysis of Fig. 11(d) and Fig. 12(d). This distinction highlights how collisional relaxation controls the response of particles to perturbations.

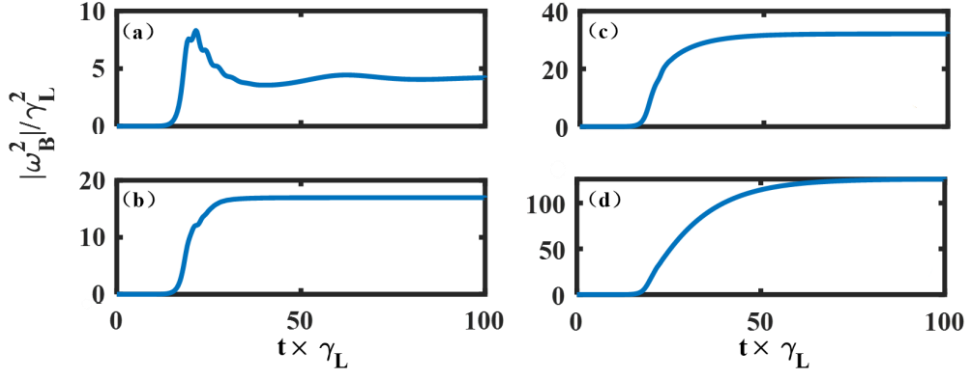

**Figure 11**. Dependences of temporal electric field evolutions on Krook coefficients, where (a-d) represents $\beta/(\gamma_L - \gamma_d)$ =0.1, 0.3, 0.5, 1, respectively. $\gamma_d/\gamma_L$ is set to 0.1437.

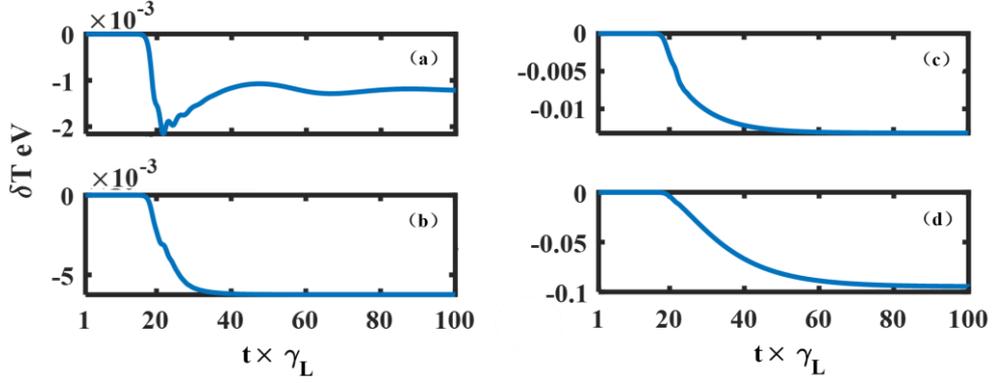

**Figure 12**. Dependences of temporal $\delta T$ evolutions on Krook coefficients, where the conditions are same as Fig. 11.

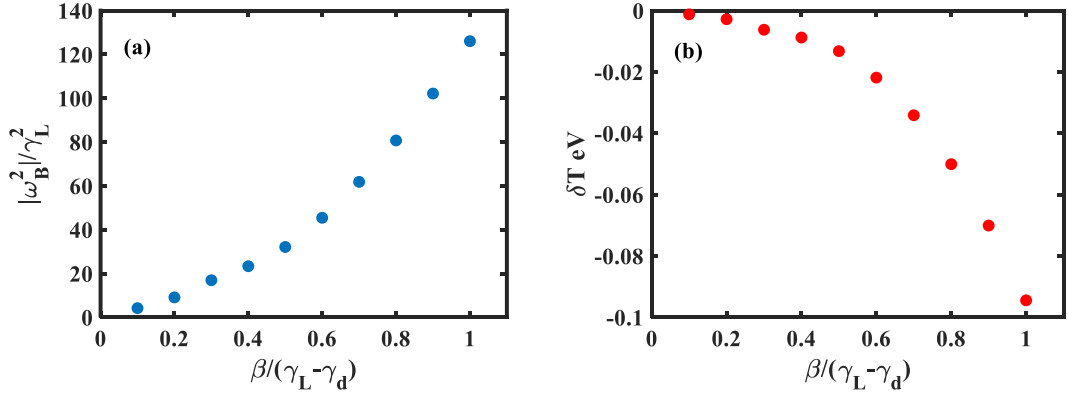

**Figure 13**. Dependences of (a) saturation time of electric field shown in Fig. 11 and (b) saturation

time of $\delta T$ shown in Fig. 12 on Krook coefficient.

Figure 14 shows 2D evolution of EEDF with four typical Krook coefficients. We can see that there are significant changes of EEDF around $t \times \gamma_L \approx 18$. With stronger Krook relaxation process, the peak is larger, e.g. from $4.5 \times 10^6$ m$^{-4}$s for $\beta/(\gamma_L - \gamma_d) = 0.1$ to $1.4 \times 10^7$ m$^{-4}$s for $\beta/(\gamma_L - \gamma_d) = 1$. The broadening of velocity-space perturbations indicates enhanced kinetic energy loss from the particle population. Additionally, the stabilization timescale of the perturbed EEDF increases with higher $\beta/(\gamma_L - \gamma_d)$, e.g. for $\beta/(\gamma_L - \gamma_d) = 0.4$, stabilization occurs at $t \times \gamma_L \approx 25$; for $\beta/(\gamma_L - \gamma_d) = 1$, stabilization requires $t \times \gamma_L \approx 60$. This temporal evolution aligns with the electric field evolution shown in Fig. 11.

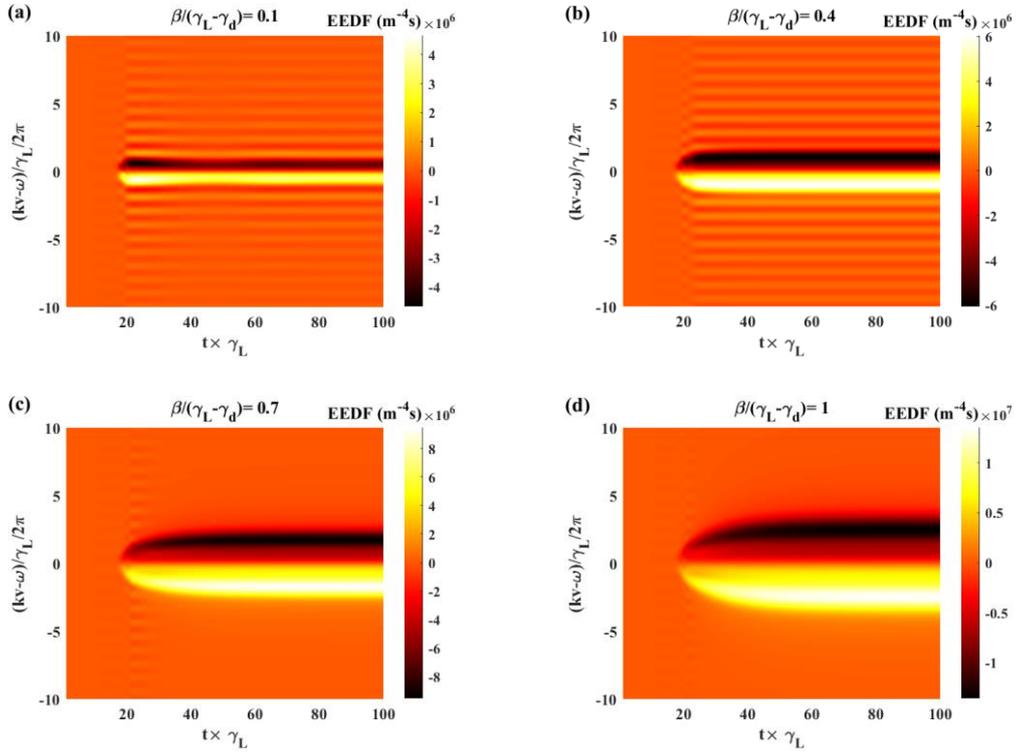

**Figure 14**. 2D evolution of disturbed EEDF with different Krook coefficients, where (a-d) represent $\beta/(\gamma_L - \gamma_d) = 0.1, 0.4, 0.7$ and 1, respectively.

Based on the above contents, a large $\beta$-range scan study is conducted, which is more instructive for helicon plasma under the antenna comparing to downstream plasma. Furthermore, there would be more energetic electrons in devices with higher power such as MPEX (Material Plasma Exposure Experiment)[52] and VASIMR(variable specific impulse magnetoplasma rocket) [53].

### 4.3 Effects of $\gamma_L$ on BOT instability

Refering the Eq. (7) and Eq. (14), Our former work assumes that the slope of the distribution function of the energetic electron population at resonance is invariant and change $\beta$ to study higher energetic eletron population when BOT instability occurs. Now, we assume that $\beta$ is invariant, and change $\gamma_L$ to study its effects on the BOT instability.

The research presented above refers to the HMHX device which has power level of 1500 W. We now consider the scenarios of much higher power, such as the MPEX and VASIMR that can be operated at >200 kW. High power usually means high temperature of background electrons and large population of energetic electrons. Both of them indicate low amplitude of $\gamma_d / \gamma_L$. In this sense, we decrease the ratio of $\gamma_d / \gamma_L$ from 0.13 (slightly lower than HMHX parameters) to 0.02 (lowest value to get stable solution), to gradually account for higher power scenarios.

It can be seen from Fig. 15 that when the ratio of $\gamma_d / \gamma_L$ decreases, the first extreme value of the electric field increases, and the largest first extreme value appears in the Fig. 15(e), that is, when the power is the largest. This can also be observed in Fig. 16(a), when $\gamma_d / \gamma_L$ is minimized ($\gamma_d / \gamma_L = 0.02$), the system exhibits the largest perturbation distribution function, indicating more intense wave-particle interactions under this condition. In addition, lower values of $\gamma_d / \gamma_L$ not only effect the maximum amplitude of the wave field but also influence its steady-state value. As shown in Fig 15(a) to (e), the steady-state value exhibits a gradually rise as decreasing $\gamma_d / \gamma_L$. Moreover, Fig. 15(a) to Fig. 15(e) demonstrate distinct relaxation rates. Specifically, Fig. 15(a) reaches the steady-state regime around $t \times \gamma_L = 40$ while Fig. 15(b) to (e) require a longer duration to attain their steady-state values.

The impact of variations in $\gamma_d / \gamma_L$ on the distribution function is also a key focus of our investigation. As revealed in Fig. 16(a), a reduction in $\gamma_d / \gamma_L$ not only amplifies perturbations but also intensifies their localization near the resonance region. For instance, when $\gamma_d / \gamma_L$ ranges from 0.13 to 0.04 within the $|(kv - \omega) / \gamma_L / 2\pi| \approx 4$ domain, $f_0$ is close to stable, whereas persistent strong perturbations are observed at $\gamma_d / \gamma_L = 0.02$. This implies that under higher-power conditions, the BOT instability exhibits an expanded spatial influence. Similarly, in Fig. 16(b), as $\gamma_d / \gamma_L$ decreases, a broader region of the distribution function becomes flattened, leading to the formation of a more extensive 'plateau'. Moreover, while $\gamma_d / \gamma_L$ significantly modulates the steady-state perturbation amplitude and spatial distribution, it do not change the instability onset time, as evidenced by Fig. 17(a)-(d). And the reduction in $\gamma_d / \gamma_L$ leads to an increase in the peak amplitude., e.g. from $4.5 \times 10^6$ m$^{-4}$s for $\gamma_d / \gamma_L = 0.13$ (Fig. 17(a)) to $6.5 \times 10^6$ m$^{-4}$s for $\gamma_d / \gamma_L = 0.02$ (Fig. 17(d)). Collectively, these observations demonstrate that high-power conditions result in significantly enhanced wave-particle interactions, leading to the development of pronounced instabilities.

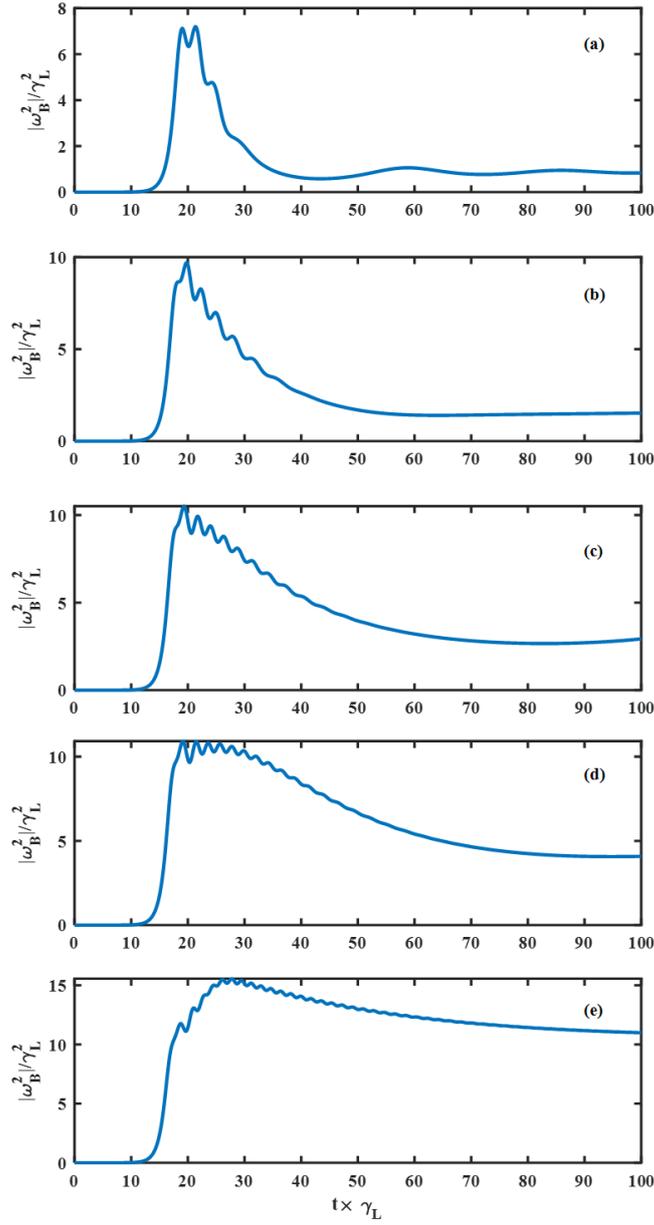

**Figure 15.** Evolution of the electric field with different ratios of $\gamma_d / \gamma_L$, where (a-e) represents $\gamma_d / \gamma_L = 0.13, 0.07, 0.05, 0.04, 0.02$, respectively.

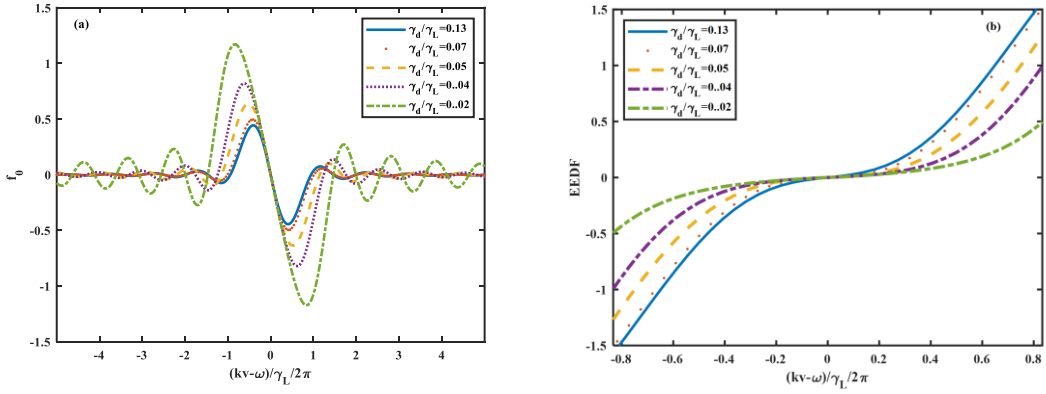

**Figure 16.** Evolution of disturb EEDF (a) and EEDF(b) with different ratios of $\gamma_d / \gamma_L$, where the conditions are same as Fig. 15.

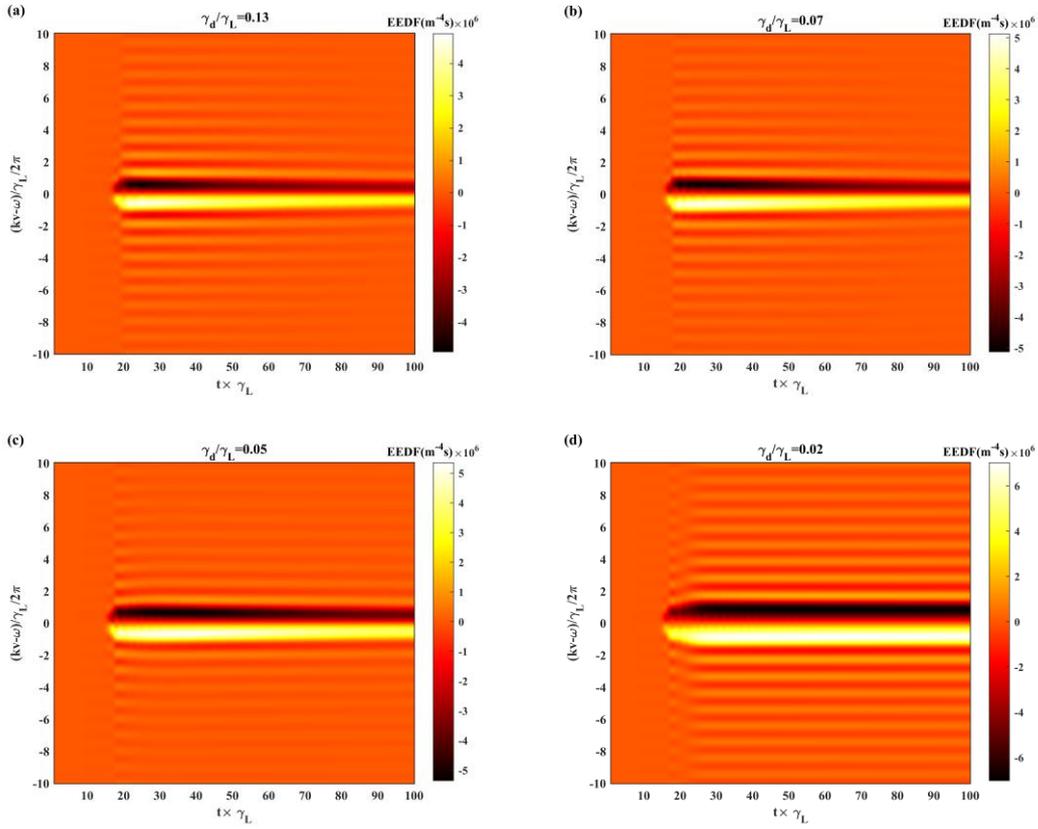

**Figure 17.** 2D evolution of disturbed EEDF with different $\gamma_d / \gamma_L$.

### 4.4 Effects of BOT instability on bulk plasma

In previous sections, the discussions are focused on the BOT instability itself. The influence of BOT instability on bulk plasma will be presented below via fluid equations and Poisson equation. The frequency of disturbed mode is much higher than the characteristic frequency of ion, therefore, ion is regarded as stationary. The equation of mass conservation for bulk electrons is:

$$\frac{\partial n_e}{\partial t} + \frac{\partial n_e u_e}{\partial x} = 0, \quad (15)$$

where electron density is $n_e(x,t) = n_0(x) + n_1(x,t)$ and electron fluid velocity is $u_e(x,t) = u_0(x) + u_1(x,t)$. The subscripts 0 and 1 denote equilibrium and perturbation, respectively. The electron momentum equation is:

$$m_e n_e \left( \frac{\partial u_e}{\partial t} + u_e \frac{\partial u_e}{\partial x} \right) = -e n_e E. \quad (16)$$

Here, electron pressure is neglected, and since Langmuir wave is driven by pressure, our attention is focused on local effect rather its propagation. In addition, it will be different from stationary plasma to rotating plasma which moves macroscopically, e.g. CSDX (controlled shear decorrelation experiment) plasma. The Poisson equation has the form of:

$$\frac{\partial E}{\partial x} = \frac{e(n_0 - n_e)}{\varepsilon_0}. \quad (17)$$

In CSDX, both plasma density and plasma rotation are radially nonuniform [12]. There is particle transport barrier caused by several macroscopic instabilities. It is interesting to know the influence of BOT instability on bulk plasma. The BOT instability is introduced by saturated electric field $E$ in Eq. (16) and Eq. (17) while bulk plasma refers to CSDX device. It makes sense because the BOT instability relates to energetic electron while only a perturbation of electric field is needed by fluid equations. In CSDX device, plasma is radially rotated, and a radial gradient of rotated velocity also exists. As a result, the BOT instability is supposed to appear in radial and bulk plasma that is radially non-uniform.

Equations (15)-(17) are linearized by Fourier form $\sim \exp(ikx - i\omega t)$, where wave comes from the BOT instability. Density perturbation is derived as:

$$n_1 = \frac{\beta_1 c - b\gamma_1 - e n_0 b E_1}{\alpha_1 b - \beta_1 a}, \quad (18)$$

*With the parameters a, b, c, $\alpha_1$, $\beta_1$ and $\gamma_1$:*

$$\begin{cases} a = -i\omega + iku_0 + \dfrac{\partial u_0}{\partial x}, \\ b = \dfrac{\partial n_0}{\partial x} + ikn_0, \\ c = u_0\dfrac{\partial n_0}{\partial x} + n_0\dfrac{\partial u_0}{\partial x}, \\ \alpha_1 = m_e u_0 \dfrac{\partial u_0}{\partial x}, \\ \beta_1 = -i\omega m_e n_0 + m_e n_0 \dfrac{\partial u_0}{\partial x} + ikm_e n_0 u_0, \\ \gamma_1 = m_e n_0 u_0 \dfrac{\partial u_0}{\partial x}. \end{cases} \quad (19)$$

For stationary plasma, the expression of perturbed density is simply:

$$n_1 = \frac{-ebE_1}{\omega^2 m_e}, \quad (20)$$

and its real part is,

$$\text{Re}(n_1) = \frac{-\omega_B^2}{\omega^2 k}\frac{\partial n_0}{\partial x}. \quad (21)$$

The calculation is located at $r = 4.5$ cm in CSDX device, where density and velocity of bulk plasma are $n_0 = 2.54\times 10^{18}$ m$^{-3}$ and $u_0 = 1.1$ km/s. The gradients of density and velocity presented in are $\partial n_0/\partial x = -3.95\times 10^{20}$ m$^{-4}$ and $\partial n_0/\partial x = 15104$ s$^{-1}$. The perturbing electric field $E_1$ in Eq. (18) from BOT instability is shown in Fig. 13(a). Figure 18 demonstrates the influence of BOT instability on bulk plasma density for stationary and rotating plasmas, respectively. It is found that the effect of BOT instability on a rotated plasma is nearly an order of magnitude greater than that on a stationary plasma. The order of perturbed density for rotated case is $\sim 10^{12}$ m$^{-3}$.

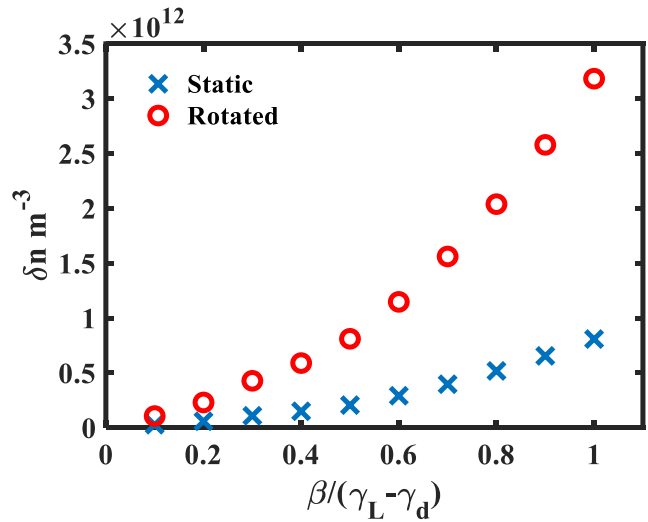

**Figure 18**. Dependence of perturbed plasma density on Krook coefficient for static and rotational

plasma, respectively.

In helicon plasma, one of fascinating phenomena is "blue core" or "bright core" discharge which can yield 100% ionization. The transport barrier relating to the formation of blue core has been detected in many devices. Fluid instability such as DW, K-H, R-T instabilities are proposed drives to form the barrier [12]. Notably, the interplay between the transport barrier and BOT instability significantly impacts the kinetics of energetic electrons and even alters the motion of bulk electrons. To quantify these effects, we calculate the corresponding forces and fluxes in the following.

As known, if there is a spatially non-uniform and high-frequency oscillating electric field, pondermotive force (Eq.(21)) would appear,

$$\vec{F} = -(e^2/4m\omega^2)\nabla E_0^2 \tag{21}$$

From Eqs. (15)-(17) and experiments in CSDX, the wave or oscillation is spatially inhomogeneous in a non-uniform plasma. Therefore, pondermotive force is calculated and the result is presented in Fig. 19(a). One can see that pondermotive force in this case is too small for ions but large for electrons. Notably, if the force acts on electron for a while, charge separation would occur. Further, electron flux, $\langle \Gamma_r \rangle = \langle n_1 u_1 \rangle$ [49] is calculated to evaluate particle transport to explore the possible transport barrier. The result in Fig. 19(b) indicates that flux also increases with Krook coefficient. The positive value reveals that electron transports outward, which means that the axial transport decreases due to mass flow conservation.

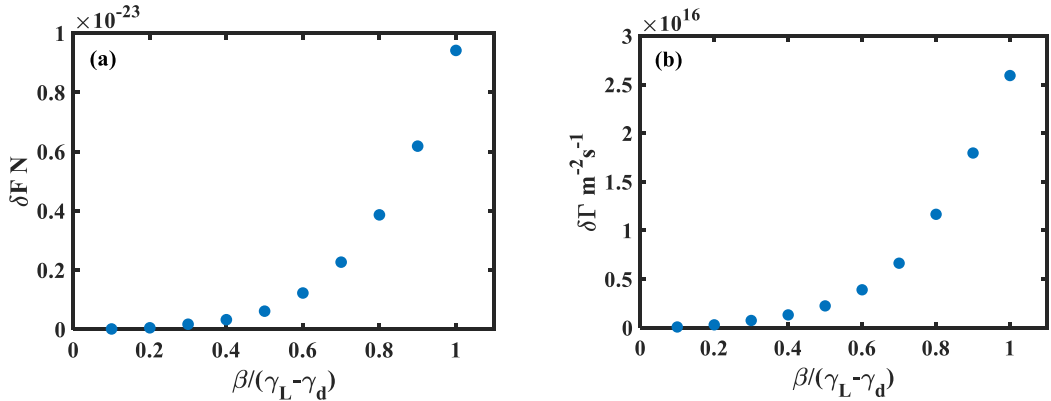

**Figure 19**. Dependence of pondermotive force (a) and electrons flux (b) on Krook coefficient.

In addition, an enlightening expectation is the influence of BOT instability on double layer in helicon source. The relevant experiment denotes that there are energetic upstream electrons with sufficient energy to overcome the potential of the double layer [19]. When there is BOT instability appearing in the upstream of double layer, the potential will be changed. The BOT instability in double layer and its effect will be studied in the future when the EEDF is available.

## 5. Conclusion

In this paper, the BOT instability excited by energetic electrons in helicon plasma is studied based on experimental data from the HMHX device. The plasma oscillation is regarded as a single mode and the main relaxation is Krook-type in the BOT instability. The influences of BOT instability on bulk plasma (in CSDX device), the effects of Krook operators on BOT instability and the influences of BOT instability for high-power helicon devices are investigated, respectively. The important findings are summarized below.

- There is BOT instability indeed existing in helicon plasma sources such as HMHX device. According to experiments in HMHX, the growth rate of wave is $1.71 \times 10^8$ s$^{-1}$, the wave damping mainly comes from electron-electron and electron-ion collision, and the ratio of wave damping to growth rate is 0.1437.

- Under the plasma parameters of HMHX, after the BOT instability is excited, the distribution function transits from explosion oscillation to stability, and the phase island appears periodically in the phase space. The disturbance is confined to the potential well, and the wave frequency does not change.

- In the calculation, the wave electric field will reach saturation for a wide range of Krook coefficients. Plasma with greater relaxation process allows the evolution of BOT instability to evolve into steady state with slower but larger saturation.

- The change in EEDF caused by BOT instability is about 2.5% of the original distribution of energetic electrons in the HMHX device. There are more energetic electrons participating in wave-particle interaction with greater Krook coefficient which is the rate of recovering the initial distribution.

- The disturbed temperature $\delta T$ is proposed to evaluate the influence of BOT instability on energetic electrons. The $\delta T$ value decreases exponentially with the enhanced Krook relaxation process.

- Higher-power plasma devices that can drive stronger BOT instability are considered as well. Results show that higher power will largely lead to getting steady state slower after the initial BOT instability excitation, and impact a broader population of electrons.

- The fluid model reveals that the BOT instability would effect the macroscopic characteristics, e.g. the density and flux of bulk plasma. The perturbation of bulk electron density for rotating plasma is an order larger than that for stationary plasma.

- When the BOT instability appears in the radial direction, bulk electrons will transport radially outwards, and then the axial transport would decrease, because of the momentum conservation. This may influence the propulsion using helicon plasma, such as helicon double-layer thruster.

Future research will be devoted to the experimental measurements of BOT instability in high-power helicon experiments and their detailed comparisons with computed results.

## Acknowledgement

This work is supported by National Natural Science Foundation of China (92271113, 12411540222, 12481540165), Fundamental Research Funds for Central Universities (2022CDJQY-003), and Chongqing Entrepreneurship and Innovation Support Program for Overseas Returnees (CX2022004).

## Data availability statement

The data that support the findings of this study are available from the corresponding authors upon reasonable request.

## References

[1] M. A. Lieberman, A.J. Lichtenberg. Principles of plasma discharges and materials processing[J]. John Wiley & Sons, Inc, 1994.

[2] D. Kuwahara, S. Shinohara, K. Yano. Thrust characteristics of high-density helicon plasma using argon and xenon gases[J]. Journal of Propulsion and Power, 2017, 33(2): 420-424.

[3] F. R. C. Diaz. The VASIMR Rocket Rockets used to be of two types: powerful but fuel-guzzling, or efficient but weak[J]. Scientific American, 2000, 283(5): 72-85.

[4] P. K. Loewenhardt, B. D. Blackwell, R.W. Boswell, et al. Plasma production in a toroidal heliac by helicon waves[J]. Physical Review Letters, 1991, 67(20): 2792.

[5] R. W. Boswell. Plasma production using a standing helicon wave[J]. Physics Letters A, 1970, 33(7): 457-458.

[6] S. Shinohara, T. Tanikawa. Characteristics of a large volume helicon plasma source[J]. Physics of Plasmas, 2005, 12(4).

[7] K. Takahashi, S. Takayama, A. Komuro, et al. Standing helicon wave induced by a rapidly bent magnetic field in plasmas[J]. Physical Review Letters, 2016, 116(13): 135001.

[8] L. Chang, M. J. Hole, J. F. Caneses, et al. Wave modeling in a cylindrical non-uniform helicon discharge[J]. Physics of Plasmas, 2012, 19(8).

[9] F. F. Chen, D. Arnush. Generalized theory of helicon waves. I. Normal modes[J]. Physics of Plasmas, 1997, 4(9): 3411-3421.

[10] M. Krämer, Y. M. Aliev, A. B. Altukhov, et al. Anomalous helicon wave absorption and parametric excitation of electrostatic fluctuations in a helicon-produced plasma[J]. Plasma Physics and Controlled Fusion, 2007, 49(5A): A167.

[11] T. Zhang, R. Cui, W. Zhu, et al. Influence of neutral depletion on blue core in argon helicon plasma[J]. Physics of Plasmas, 2021, 28(7).

[12] S. C. Thakur, C. Brandt, L. Cui, et al. Multi-instability plasma dynamics during the route to

fully developed turbulence in a helicon plasma[J]. Plasma Sources Science and Technology, 2014, 23(4): 044006.

[13] F. F. Chen. Plasma ionization by helicon waves[J]. Plasma Physics and Controlled Fusion, 1991, 33(4): 339.

[14] A. R. Ellingboe, R. W. Boswell, J.P. Booth, et al. Electron beam pulses produced by helicon-wave excitation[J]. Physics of Plasmas, 1995, 2(6): 1807-1809.

[15] A. W. Molvik, A. R. Ellingboe, T.D. Rognlien. Hot-electron production and wave structure in a helicon plasma source[J]. Physical Review Letters, 1997, 79(2): 233.

[16] R. T. S. Chen, N. Hershkowitz. Multiple electron beams generated by a helicon plasma discharge[J]. Physical Review Letters, 1998, 80(21): 4677.

[17] K. Takahashi, H. Akahoshi, C. Charles, et al. High temperature electrons exhausted from rf plasma sources along a magnetic nozzle[J]. Physics of Plasmas, 2017, 24(8).

[18] S. Ghosh, S. Yadav, K.K. Barada, et al. Formation of annular plasma downstream by magnetic aperture in the helicon experimental device[J]. Physics of Plasmas, 2017, 24(2).

[19] K. Takahashi, C. Charles, R. Boswell, et al. Transport of energetic electrons in a magnetically expanding helicon double layer plasma[J]. Applied Physics Letters, 2009, 94(19).

[20] P. K. Loewenhardt, B. D. Blackwell, S. M. Hamberger. Production of fast electrons in a prototype heliac by helicon waves[J]. Physics of Plasmas, 1994, 1(4): 875-880.

[21] B. B. Sahu, A. Ganguli, R. D. Tarey. Warm electrons are responsible for helicon plasma production[J]. Plasma Sources Science and Technology, 2014, 23(6): 065050.

[22] J. L. Kline, E. E. Scime, R. F. Boivin, et al. RF absorption and ion heating in helicon sources[J]. Physical Review Letters, 2002, 88(19): 195002.

[23] S. A. Cohen, X. Sun, N.M. Ferraro, et al. On collisionless ion and electron populations in the magnetic nozzle experiment (MNX)[J]. IEEE Transactions on Plasma Science, 2006, 34(3): 792-803.

[24] L. Chang, M. J. Hole, C. S. Corr. A flowing plasma model to describe drift waves in a cylindrical helicon discharge[J]. Physics of Plasmas, 2011, 18(4).

[25] H. L. Berk, B. N. Breizman, M. Pekker. Nonlinear dynamics of a driven mode near marginal stability[J]. Physical Review Letters, 1996, 76(8): 1256.

[26] G. Zhang, T. Huang, J. Chenggang, et al. Development of a helicon-wave excited plasma facility with high magnetic field for plasma–wall interactions studies[J]. Plasma Science and Technology, 2018, 20(8): 085603.

[27] F. F. Chen, D. D. Blackwell. Upper limit to Landau damping in helicon discharges[J]. Physical Review Letters, 1999, 82(13): 2677.

[28] M. U. Siddiqui, J. S. McKee, J. McIlvain, et al. Electron heating and density production in microwave-assisted helicon plasmas[J]. Plasma Sources Science and Technology, 2015, 24(3): 034016.

[29] I. B. Bernstein, J. M. Greene, M.D. Kruskal. Exact nonlinear plasma oscillations[J]. Physical Review, 1957, 108(3): 546.

[30] H. L. Berk, B. N. Breizman. Saturation of a single mode driven by an energetic injected beam. I. Plasma wave problem[J]. Physics of Fluids B: Plasma Physics, 1990, 2(9): 2226.

[31] H. L. Berk, B. N. Breizman. Saturation of a single mode driven by an energetic injected beam. II. Electrostatic "universal" destabilization mechanism[J]. Physics of Fluids B: Plasma Physics, 1990, 2(9): 2235-2245.

[32] H. L. Berk, B. N. Breizman. Saturation of a single mode driven by an energetic injected beam. III. Alfvén wave problem[J]. Physics of Fluids B: Plasma Physics, 1990, 2(9): 2246.

[33] H. L. Berk, B. N. Breizman, H. Ye. Scenarios for the nonlinear evolution of alpha-particle-induced Alfvén wave instability[J]. Physical Review Letters, 1992, 68(24): 3563.

[34] B. N. Breizman, H. L. Berk, H. Ye. Collective transport of alpha particles due to Alfvén wave instability[J]. Physics of Fluids B: Plasma Physics, 1993, 5(9): 3217-3226.

[35] H. L. Berk, B. N. Breizman, M. Pekker. Numerical simulation of bump-on-tail instability with source and sink[J]. Physics of Plasmas, 1995, 2(8): 3007-3016.

[36] M. K. Lilley, B. N. Breizman, S. E. Sharapov. Destabilizing effect of dynamical friction on fast-particle-driven waves in a near-threshold nonlinear regime[J]. Physical Review Letters, 2009, 102(19): 195003.

[37] K. L. Wong, R. Majeski, M. Petrov, et al. Evolution of toroidal Alfvén eigenmode instability in tokamak fusion test reactor[J]. Physics of Plasmas, 1997, 4(2): 393-404.

[38] S. D. Pinches, H. L. Berk, M. P. Gryaznevich, et al. Spectroscopic determination of the internal amplitude of frequency sweeping TAE[J]. Plasma Physics and Controlled Fusion, 2004, 46(7): S47.

[39] M. P. Gryaznevich, S. E. Sharapov. Perturbative and non-perturbative modes in START and MAST[J]. Nuclear Fusion, 2006, 46(10): S942.

[40] B. N. Breizman, H. L. Berk, M. S. Pekker, et al. Critical nonlinear phenomena for kinetic instabilities near threshold[J]. Physics of Plasmas, 1997, 4(5): 1559-1568.

[41] M. K. Lilley, B. N. Breizman, S. E. Sharapov. Effect of dynamical friction on nonlinear energetic particle modes[J]. Physics of Plasmas, 2010, 17(9).

[42] M. K. Lilley, B. N. Breizman. Convective transport of fast particles in dissipative plasmas near an instability threshold[J]. Nuclear Fusion, 2012, 52(9): 094002.

[43] R. M. Nyqvist, M. K. Lilley, B. N. Breizman. Adiabatic description of long range frequency sweeping[J]. Nuclear Fusion, 2012, 52(9): 094020.

[44] M. K. Lilley, R. M. Nyqvist. Formation of phase space holes and clumps[J]. Physical Review Letters, 2014, 112(15): 155002.

[45] F. Eriksson, R. M. Nyqvist, M. K. Lilley. Kinetic theory of phase space plateaux in a non-thermal energetic particle distribution[J]. Physics of Plasmas, 2015, 22(9).

[46] R. F. Heeter, A.F. Fasoli, S. E. Sharapov. Chaotic regime of Alfvén eigenmode wave-particle

interaction[J]. Physical Review Letters, 2000, 85(15): 3177.

[47] M. Lesur, Y. Idomura. Nonlinear categorization of the energetic-beam-driven instability with drag and diffusion[J]. Nuclear Fusion, 2012, 52(9): 094004.

[48] M. Lesur, Y. Idomura, X. Garbet. Fully nonlinear features of the energetic beam-driven instability[J]. Physics of Plasmas, 2009, 16(9).

[49] R. He, Y. Xiaoyi, X. Chijie, et al. Experimental observation of the transport induced by ion Bernstein waves near the separatrix of magnetic nulls[J]. Plasma Science and Technology, 2022, 24(11): 115001.

[50] https://github.com/mklilley/BOT

[51] D. Jing, L. Chang, X. Yang, et al. Exploration on the possible bump-on-tail instability in VASIMR[J]. Space: Science & Technology, 2024, 4: 0107.

[52] J. Rapp, A. Lumsdaine, C.J. Beers, et al. Latest results from Proto-MPEX and the future plans for MPEX[J]. Fusion Science and Technology, 2019, 75(7): 654-663.

[53] F. R. C. Diaz. The VASIMR rocket[J]. Scientific American, 2000, 283(5): 90-97.